%% file: preprint-template.tex
\documentclass[twocolumn, switch]{article} 

\usepackage{preprint}

\usepackage{amsmath, amsthm, amssymb, amsfonts}

\usepackage[numbers,square]{natbib}
\usepackage[utf8]{inputenc}	
\usepackage[T1]{fontenc}	
\usepackage{xcolor}		
\usepackage[plainpages=false,
    pdfpagelabels,
    bookmarksnumbered,
    colorlinks=false,
    hidelinks,
    breaklinks,
    unicode,
    pdfencoding=auto,
    bookmarks=true,
    pdfstartview=Fit]{hyperref}
\usepackage{booktabs} 		
\usepackage{nicefrac}		
\usepackage{microtype}		
\usepackage{lineno}		
\usepackage{float}			

\usepackage{glossaries}

\usepackage{newfloat}
\DeclareFloatingEnvironment[name={Supplementary Figure}]{suppfigure}
\usepackage{sidecap}
\sidecaptionvpos{figure}{c}

\usepackage{titlesec}
\titlespacing\section{0pt}{12pt plus 3pt minus 3pt}{1pt plus 1pt minus 1pt}
\titlespacing\subsection{0pt}{10pt plus 3pt minus 3pt}{1pt plus 1pt minus 1pt}
\titlespacing\subsubsection{0pt}{8pt plus 3pt minus 3pt}{1pt plus 1pt minus 1pt}

\usepackage{tikz,xcolor,hyperref}

\definecolor{lime}{HTML}{A6CE39}
\DeclareRobustCommand{\orcidicon}{
	\begin{tikzpicture}
	\draw[lime, fill=lime] (0,0) 
	circle [radius=0.16] 
	node[white] {{\fontfamily{qag}\selectfont \tiny ID}};
	\draw[white, fill=white] (-0.0625,0.095) 
	circle [radius=0.007];
	\end{tikzpicture}
	\hspace{-2mm}
}
\foreach \x in {A, ..., Z}{\expandafter\xdef\csname orcid\x\endcsname{\noexpand\href{https://orcid.org/\csname orcidauthor\x\endcsname}
			{\noexpand\orcidicon}}
}

\title{Defect-Aware Physics-Based Compact Model for Ferroelectric nvCap: From TCAD Calibration to Circuit Co-Design}

\usepackage{xwatermark}
\newwatermark[firstpage,color=gray!90,angle=0,scale=0.28, xpos=0in,ypos=-5in]{*correspondence: \texttt{luca.fehlings@tum.de}}

\usepackage{authblk}

\author[1,2\thanks{\tt{luca.fehlings@tum.de}}]{Luca Fehlings\orcidA{}}
\author[3]{Nihal Raut\orcidB{}}
\author[3]{Md. Hanif Ali\orcidC{}}
\author[4]{Francesco Maria Puglisi\orcidD{}}
\author[5]{Andrea Padovani\orcidE{}}
\author[3]{Veeresh Deshpande\orcidF{}}
\author[1,2]{Erika Covi\orcidG{}}

\affil[1]{Technical University of Munich, Germany; TUM School of Computation, Information and Technology, Department of Computer Engineering}
\affil[2]{Zernike Institute for Advanced Materials, University of Groningen, Netherlands}
\affil[3]{Department of Electrical Engineering, Indian Institute of
Technology Bombay, India}
\affil[4]{Department of Engineering “Enzo Ferrari”, University of Modena and Reggio Emilia, Italy}
\affil[5]{Department of Engineering Sciences and Methods, University of Modena and Reggio Emilia, Italy.}

\begin{document}

\input{glossary}

\twocolumn[ 
  \begin{@twocolumnfalse} 
  
\maketitle

\begin{abstract}
Ferroelectric non-volatile capacitance-based memories enable non-destructive readout and low-power in-memory computing with 3D stacking potential. However, their limited memory window ($1\text{--}10\,\text{fF}/\text{\textmu m}^2$) requires material-device-circuit co-optimization. Existing compact models fail to capture the physics of small-signal capacitance, device variability, and cycling degradation, which are critical parameters for circuit design.
In non-volatile capacitance devices, the small-signal capacitance difference of the polarization states is the key metric. The majority of the reported compact models do not incorporate any physical model of the capacitance as a function of voltage and polarization. 
We present a physics-based compact model that captures small-signal capacitance, interface and bulk defect contributions, and device variations through multi-scale modeling combining experimental data, TCAD simulations, and circuit validation.
Based on this methodology, we show optimized memory read-out with $\pm$5\,mV sense margin and impact of device endurance at the circuit level. This work presents a comprehensive compact model which enables the design of selector-less arrays and 3D-stacked memories for compute-in-memory and storage memory.
\end{abstract}
\vspace{0.35cm}

  \end{@twocolumnfalse} 
] 



\section{Introduction}
\label{sec:introduction}

\begin{figure}[t]
        \centering
        \includegraphics[]{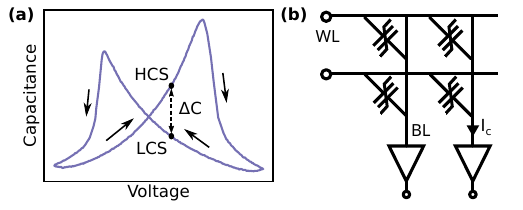}
        \caption{\textbf{(a)} Illustration of the nvCap capacitance and polarization hysteresis. A capacitive memory window $\Delta C$ can be extracted as difference between the \gls{hcs} and the \gls{lcs}. \textbf{(b)} Possible application in a selector-less crossbar array enabled by the capacitive nature of the memory device.}
        \label{fig:iedm25:1}
    \end{figure}

    Ferroelectric technology has emerged as a promising low-power alternative to conventional \gls{cmos} solutions for storage memory and \gls{cim} applications. Recent research focuses on exploring non-destructive read methodologies through \gls{nvcap} devices. The built-in electric field in the ferroelectric capacitor results in the presence of a non-volatile small-signal capacitance that depends on the polarization of the ferroelectric, determining a \gls{hcs} and \gls{lcs} \cite{iedm23_capreadwindow, phadke2024reliability}. This phenomenon, often modeled as a polarization-dependent dielectric constant, facilitates the non-destructive read functionality. While electronic mechanisms contribute to this capacitance hysteresis, as demonstrated by GHz-regime impedance measurements \cite{kim_iedm2023}, potentially enabling high-speed read operations, also slower domain-wall dynamics have shown to contribute to this phenomenon \cite{koduru2025small, pevsic2016physical}. Further, the capacitance can be read with voltages well below the coercive voltage of the device and a low read-disturb \cite{fehlings2025reliability}, while maintaining the low-voltage program and erase operations of a \gls{fecap}. The \gls{nvcap} device offers the possibility of selector-less or \gls{1t1c} memory arrays and 3D stacking \cite{iedm23_micron} for high-density memory (Fig.~\ref{fig:iedm25:1}). However, \gls{nvcap} devices have a limited memory window (1-10\,fF/\textmu m²) and the read signal is severely limited by the bit line capacitance \cite{xiang2024compact}. While material stacks optimized for \gls{nvcap} operation are already being researched \cite{huang2025large}, there is a need for careful co-design of the material stack, devices, and memory circuit to ensure reliability and performance. Therefore, it is necessary to develop a physics-based compact model that incorporates the key material and device physics to allow reliable circuit design as well as Monte Carlo, post-layout and noise verification. Further, critical device and material parameters need to be identified to enable \gls{dtco} to overcome the low signal amplitude particularly for high bit line capacitances in large arrays. Conventional ferroelectric capacitor models \cite{tung-2021,yan-2023,pesic-2017,kim-2020,alessandri-2018,feng-2022,lederer-2023} cannot support design flows that require device-circuit co-optimization with reliability, aging, noise, and mismatch. These models have three critical limitations: (1) they lack self-consistent and physics-based small-signal capacitance modeling, (2) they ignore device-to-device variations, (3) they ignore defects and related degradation mechanisms and (4) they cannot be calibrated with TCAD simulations. \Gls{tcad} simulations are critical for the verification and the calibration of compact models, in particular for parameters that are difficult to observe directly, such as defects. As a result of these shortcomings, the design and optimization of the circuits needed to read and program these devices is challenging. In this study, we propose a \gls{dtco} framework based on experimental data, \gls{tcad} simulations, and a VerilogA compact model. This approach extends an existing compact model by a interface-defect-based capacitance and a bulk-defect-based leakage current model. Therefore, the model can be verified by \gls{tcad} simulations in addition to experimental data, and \gls{dtco} can be performed by providing a comprehensive link between the extracted defect parameters and the electrical behavior. Based on this, we present a read-circuit that produces a 10~mV memory window at the bit line, showing the need for sub-mV offset-calibrated sense amplifiers and aggressive bit line capacitance reduction for reliable and degradation-aware memory designs.

\section{Device Fabrication and Characterization}
    We demonstrate our framework using W-Al\textsubscript{2}O\textsubscript{3}-Hf\textsubscript{0.5}Zr\textsubscript{0.5}O\textsubscript{2}-W bi-layer capacitors (Fig.~2). The presence of asymmetric interfaces gives rise to an in-built electric field in the device at 0 bias, leading to a non-volatile capacitance. The stack (Fig.~\ref{fig:iedm25:2}(a)) comprises a 50\,nm sputtered W bottom electrode, 2\,nm Al\textsubscript{2}O\textsubscript{3}, and 10\,nm HZO layers (ALD at 200$^{\circ}$\,C), and a 50\,nm sputtered W top electrode. The device measures 100\,\textmu m\,$\times$\,100\,\textmu m. Post-metallization annealing was carried out at 450$^{\circ}$C for 720\,s in a nitrogen (N\textsubscript{2}) environment.
    The \gls{nvcap} devices are woken up with 1\,k cycles at 1\,kHz. Small signal \gls{cv} measurements are performed with quasi-static DC voltage sweep combined with small AC signal of 30\,mV amplitude.
    
\section{Material-Device-Circuit Co-design Methodology}

\begin{figure}[t]
        \centering
        \includegraphics[]{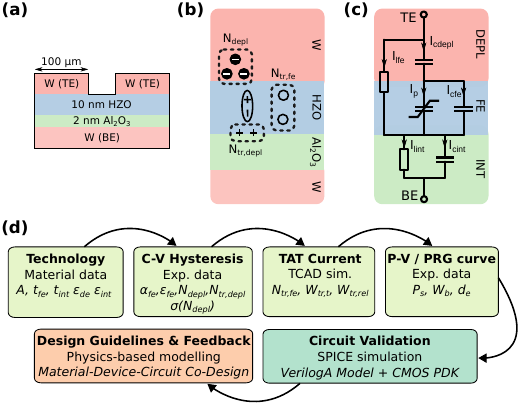}
        \caption{\textbf{(a)} Material stack of the fabricated nvCap devices. \textbf{(b)} The nvCap device behavior is impacted by non-idealities, such as the carrier density at the electrode interface leading to finite screening, bulk traps in the HZO causing leakage currents and interface traps that screen the polarization, circumventing depolarization effects from finite screening due to the electrode depletion and interface layer. (c) Equivalent circuit that models the ferroelectric switching and its interaction with the depletion effect, interface layer and leakage currents self-consistently. (d) Calibration flow with the parameters extracted in each step. The calibration is based on technology data, electrical characterization data and TCAD simulations, leading to a physics-based VerilogA compact model that enables material-device-circuit co-design by circuit validation and design feedback.}
        \label{fig:iedm25:2}
    \end{figure}

    We demonstrate a co-design methodology integrating experimental data, TCAD simulations for defect density extraction, a physics-based compact model, and read circuit validation with both the nvCap compact model and a CMOS PDK. The experimental data from a bilayer stack \gls{nvcap} device (Fig.~\ref{fig:iedm25:2}(a) are used to calibrate the compact model for ferroelectric and interface layer parameters. Furthermore, the leakage current data is reproduced in the Ginestra\texttrademark{} semiconductor device simulation \gls{tcad} ~\cite{ginestra} to characterize the behavior and extract trap properties. We then incorporate these trap properties (Fig.~\ref{fig:iedm25:2}(b)) in the compact model (Fig.~\ref{fig:iedm25:2}(c)). This includes the trap density $N_\text{tr,fe}$ for the bulk leakage current, based on the \gls{tat} conduction mechanism. The interface trap properties $N_{\text{tr,depl}\uparrow/\downarrow}$ are extrapolated from this bulk density and are calibrated against the experimental CV hysteresis. The effects of these interface defects are treated in the compact model as contributions to polarization screening and are therefore grouped together at the interface. However, they do not necessarily represent defects located exactly at the material interface; defects throughout the material stack can influence polarization screening and bias behavior.
    
    A schematic of the segments and physics included in the compact model for \gls{nvcap} is presented in Fig.~\ref{fig:iedm25:2}(b-c). The ferroelectric device model is based on our previous work ~\cite{fehlings2024heracles}, which calculates a self-consistent solution based on contributions from the ferroelectric, interface and electrode layers. In order to provide an accurate description of the non-volatile small-signal capacitance, it is necessary to incorporate interface and bulk defects into the model. In this model (Fig.~\ref{fig:iedm25:2}(b)), we consider the contribution of finite screening charges $N_\text{depl}$ in the electrode interface, bulk defects $N_\text{tr,fe}$ within the ferroelectric for leakage currents, static interface trap charges $N_\text{tr,depl}$ that contribute to the capacitance hysteresis and dynamic interface charges $(I_\text{lfe}-I_\text{lint}) \text{dt}$ that contribute to the ferroelectric polarization screening.

    Therefore, we present a compact model that includes the reliability of aging of ferroelectric devices, enabling the optimization of circuits for predictive design through a multiscale modeling approach. We model leakage currents through trap-assisted tunneling (TAT) \cite{larcher2012leakage}. This VerilogA implementation integrates with our existing device model and assumes single-tunnel events for thin films:
    \begin{equation}      
        J_\text{tat} = \frac{ q \cdot N_\text{tr,fe} \cdot x_\text{c}}{2 \tau_\text{c}},
        \label{eq:tat0}
    \end{equation}
    where $N_\text{tr,fe}$ is the bulk trap density in the ferroelectric, $x_\text{c}$ is the characteristic length where the tunneling current is maximized and $\tau_\text{c}$ is the time constant of the capture process. In this approximation, the current density is linear with the trap density, which is plausible for thin films with a single tunnel event. For films on the order of 10\,nm and thicker, where the \gls{tat} mechanism may involve multiple consecutive tunneling events for a charge to traverse the insulator, causing the current density to be super-linear with the defect density. The voltage-dependence of the current is given by the capture time constant calculated as in \cite{larcher2012leakage} in the framework of the multi-phonon theory \cite{vandelli2011physical}:
    \begin{equation}
        \tau_c^{-1} = c_0 N_\text{c} \text{exp}(-\frac{x_\text{tm}}{\lambda_\text{c}}) \text{exp}(-\frac{E_\text{c,j}}{k_\text{B} T}),
        \label{eq:tat1}
    \end{equation}
    with trap-specific constant $c_0$, conduction band density of states $N_\text{c}$, characteristic tunneling length for the capture process $\lambda_\text{c}$, thermal activation barrier for the capture process $E_\text{c,j}$, Boltzmann constant $k_\text{B}$ and temperature $T$. The characteristic distance $x_\text{tm}$ represents the trap position where the tunneling current maximizes. This position depends on the applied voltage, film thickness, and trap energy level relative to the conduction band.
    \begin{figure}[t]
        \centering
        \includegraphics[]{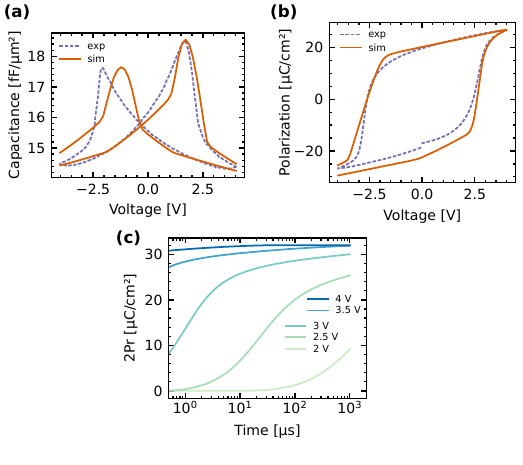}
        \caption{\textbf{(a)} Capacitance vs Voltage curve using quasi-static voltage sweep after 1000 wake-up cycles and compact model fit with the experimental curve. \textbf{(b)} Polarization vs Voltage curve after 1000 wake-up cycles using PUND sequence of 1\,kHz and fit of the compact model with the experimental curve. \textbf{(c)} Compact model simulation of the switching kinetics towards the
        down polarization state showing the transient programming behavior.}
        \label{fig:iedm25:3}
    \end{figure}
    \begin{equation}      
        x_\text{tm} = \frac{A \cdot \left(\sqrt{\frac{B \cdot V_\text{fe}^2}{V_\text{t}} + \frac{C \cdot V_\text{fe}}{V_\text{t}} + 4 t_\text{fe}^2} - 2 A\\
        t_\text{fe} + D V_\text{fe}\right)}{\lambda_\text{c} \cdot Q  V_\text{fe}^2},
        \label{eq:tat2}
    \end{equation}
    \begin{equation}
        A = 2 \cdot E_\text{tr,rel} \cdot t_\text{fe} \cdot k_\text{B} \cdot T,
        \label{eq:tat3}
    \end{equation}
    \begin{equation}   
        B = \frac{\lambda \cdot t_\text{fe}}{E_\text{tr,rel}} + \frac{\lambda_\text{c}^2}{4 V_\text{t}},
        \label{eq:tat4}
    \end{equation}
    \begin{equation}      
        C = \frac{\lambda_\text{c} \cdot t_\text{fe} \cdot (E_\text{tr,t} - E_\text{tr,rel} - \phi_\text{b,fe})}{E_\text{tr,rel}}, 
        \label{eq:tat5}
    \end{equation}
    \begin{equation}     
        D = \lambda_\text{c} \cdot t_\text{fe} \cdot q \cdot (\phi_\text{b,fe} - E_\text{tr,t} + E_\text{tr,rel}), 
        \label{eq:tat6}
    \end{equation}
    with applied voltage $V_\text{fe}$, thermal voltage $V_\text{t}$, layer thickness $t_\text{fe}$, elementary charge $q$, trap's relaxation $E_\text{tr,rel}$, and thermal ionization $E_\text{tr,t}$ energies, cathode to insulator barrier $\phi_\text{b,fe}$. The characteristic tunnel length is given by
    \begin{equation}
        \lambda_\text{c} = 2 \sqrt{\phi_\text{b,fe} 2 m^\text{*} q / \hbar^2},
        \label{eq:tat7}
    \end{equation}
    with the electron effective mass in the insulator $m^*$ and reduced Planck constant $\hbar$. The thermal activation barrier for the multi-phonon capture process is calculated via
    \begin{equation}
        E_\text{c,j} = (E_\text{tr,rel} - \Delta E_\text{j})^2/(4 E_\text{tr,rel}),
        \label{eq:tat8}
    \end{equation}
    \begin{equation}
        \Delta E_\text{j} = E_\text{tr,t} + V_\text{fe} x_\text{tm} / t_\text{fe} - \phi_\text{b,fe}.
        \label{eq:tat9}
    \end{equation}
    Elaborations on and the derivation of this \gls{tat} compact model can be found in the original publication \cite{larcher2012leakage}.
    \begin{figure}[t]
        \centering
        \includegraphics[]{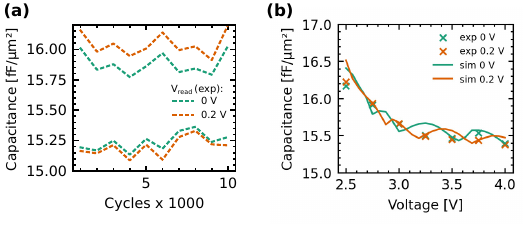}
        \caption{\textbf{(a)} Capacitance vs Cycles measured using quasi-static voltage sweep from 0\,V to 200\,mV after every 1000 cycles. \textbf{(b)} Erasing obtained by applying incrementally positive polarity pulse amplitudes, capacitance measured at DC 0\,V and 0.2\,V, and fit of the compact model with the experimental curve.}
        \label{fig:iedm25:4}
    \end{figure}

    We model the capacitance hysteresis using a voltage-dependent series capacitance that accounts for incomplete screening at the metal electrodes. As the metal electrodes are oxidized during deposition \cite{alcala2023electrode, baumgarten2024smart}, we assume electrodes with a limited charge carrier density and depletion behavior, similar to the poly-silicon depletion effect in MOS stacks.
    The depletion capacitance can be derived as
    \begin{equation}
        C_\text{depl} = \frac{\varepsilon_\text{depl}}{w_\text{depl}},
        \label{eq:cdeplbase}
    \end{equation}
    with permittivity $\varepsilon_\text{depl}$ and depletion width $w_\text{depl}$. The depletion width is defined by 
    \begin{equation}
        w_\text{depl} = \frac{D}{q N_\text{depl}},
        \label{eq:wdepl}
    \end{equation}
    with displacement field $D$, elementary charge $q$ and charge carrier density in the electrode $N_\text{depl}$. The displacement field $D_{\downarrow/\uparrow}$ is calculated separately for upwards and downwards-polarized domains. In the electrode, the displacement field has contributions from the linear displacement in the ferroelectric layer, the polarization in the ferroelectric layer $P_\text{s}$ and the charge density at the interface $qN_{\text{tr,depl}\downarrow/\uparrow}$, which originates from interface trap states:
    \begin{equation}
        D_{\downarrow/\uparrow} = \varepsilon_\text{fe} E_\text{fe} \pm (P_\text{s} - qN_{\text{tr,depl}\downarrow/\uparrow}).
        \label{eq:dupdown}
    \end{equation}
    This capacitance depends on both on the applied voltage and the polarization state.
    In accordance with the displacement field, the capacitance is calculated independently for the up- and down-polarized domains by combining Eqs. \eqref{eq:cdeplbase} -- \eqref{eq:dupdown}:
    \begin{equation}
        C_{\text{depl} \downarrow/\uparrow} = \frac{\varepsilon_\text{depl} q N_\text{depl}}{\varepsilon_\text{fe} E_\text{fe}  \pm (P_\text{s} - q N_{\text{tr,depl} \downarrow/\uparrow})}.
        \label{eq:cdepl_par}
    \end{equation}  
    The total capacitance is then proportional to the fraction of up- and down-polarized domains:
    \begin{equation}
        C_\text{depl} = p \cdot C_{\text{depl} \downarrow} + (1-p) \cdot C_{\text{depl} \uparrow},
        \label{eq:cdepl}
    \end{equation}
    with polarization $p$ normalized between 0 (up) and 1 (down). This models two parallel capacitors $C_{\text{depl} \downarrow/\uparrow}$ which are voltage-dependent, and where each contribution to the total capacitance is dependent on the polarization. The characteristic peaks in the capacitance hysteresis then arise from the gradual transition from one capacitance to the other, where both capacitances have opposing voltage dependencies. In addition, the model considers non-polar phases within the \gls{hzo} layer, i.e. a monoclinic phase portion, that are modeled as a capacitance in parallel to that of the polar, ferroelectric contributions:
    \begin{equation}
        C_\text{tot} = \alpha_\text{polar} C_\text{polar} + (1-\alpha_{polar}) C_\text{de}
    \end{equation}
    The relative phase portion $\alpha_\text{polar}$ of the ferroelectric phase and the permittivity $\varepsilon_\text{de}$ of the dielectric phase then shift the observed capacitance hysteresis up or down, while leaving the qualitative shape unchanged.
    
   The experimentally calibrated compact model is then utilized for circuit simulations along with commercial foundry \gls{cmos} \gls{pdk} to show its application in realistic circuit design.

\section{Parameter Extraction \& Model Calibration}

    \begin{figure}[t]
        \centering
        \includegraphics[]{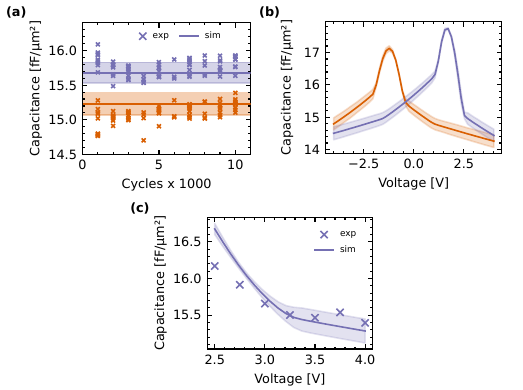}
        \caption{\textbf{(a)} Capacitance measured at 0\,V over cycles (crosses) and compact model Monte Carlo simulation of the LCS and HCS capacitance with mean and 3\,$\mathrm{\sigma}$. \textbf{(b)} Compact model Monte Carlo simulation of the capacitance hysteresis with mean and 3\,$\mathrm{\sigma}$ indicated. \textbf{(c)} Programming curve of the HCS, measured at 0\,V, and the respective compact model Monte Carlo simulation with mean and 3\,$\mathrm{\sigma}$.}
        \label{fig:iedm25:5}
    \end{figure}
    
    In our framework, we follow a hierarchical parameter extraction and model calibration procedure (Fig.~\ref{fig:iedm25:2}(d)) that draws from experimental characterization, \gls{tcad} simulations and literature on material characterization and simulation. Geometric parameters, such as the area and the layer thicknesses are based on the physical characterization or design of the device and material deposition processes. This also extends to material parameters such as the dielectric constants of the \gls{hfo2} and \gls{al2o3} layers, where small variations can be imposed to fit the model. Then, parameters are extracted from the C-V hysteresis. The base capacitance, reflected by the linear capacitors $I_\text{cfe}$ and $I_\text{cint}$, is governed by the dielectric constants of the materials $\varepsilon_\text{fe}$, $\varepsilon_\text{de}$ and $\varepsilon_\text{int}$, as well as the fraction of polar domains in the ferroelectric $\alpha_\text{polar}$. The capacitance peak is then governed by the non-linear depletion capacitance $I_\text{cdepl}$, where the carrier density $N_\text{depl}$ modulates the amplitude of the peak. The interface trap charges $N_{\text{tr,depl} \downarrow/\uparrow}$ then result in an asymmetry in the peak heights and the built-in bias, where a higher difference between the trap densities for the two polarities leads to a higher peak asymmetry and built-in bias. This also translates to the built-in bias seen in the P-V hysteresis. Then, the parameters of the tunneling/leakage currents are extracted from quasi-static leakage current measurements. Here, \gls{tcad} simulations that reproduce the leakage currents are employed to extract trap properties, such as the trapping site density $N_\text{tr,fe}$ and the trap energy levels $W_\text{tr,t}$ and $W_\text{tr,rel}$, to ensure physicality and generality of the parameters. For the calibration of the switching parameters we refer to the compact model that this framework is based on \cite{fehlings2024heracles}, where the switching model and the calibration of its parameters are explained in detail.
    For any of these steps in the calibration process, the parameters can be assigned a distribution in addition to the mean value to model mismatch or process variations. The parameters extracted using the framework for the compact model are listed in Table~\ref{tab:parameters}.

\section{Results and Discussion}
    
    \begin{figure}[t]
        \centering
        \includegraphics[]{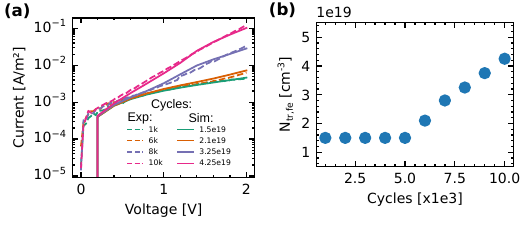}
        \caption{\textbf{(a)} Comparison of the simulated and experimental curve of leakage vs cycles to estimate the bulk defect density $N_\text{tr,fe}$ using Ginestra. The leakage current < 500\,mV is only minimally affected by increasing defects. \textbf{(b)} Bulk defect densities in the ferroelectric $N_\text{tr,fe}$ extracted from the Ginestra fitting. For low cycles the extracted defect density is constant as the leakage current is dominated by resistive current.}
        \label{fig:iedm25:6}
    \end{figure}
    
    Our \gls{spice} compact model uses a polarization switching mechanism modulated by depolarization fields from electrode depletion and interface layers \cite{fehlings2024heracles}. An initial calibration of the model is carried out by extracting its parameters from experimental electrical characterization data. For this, the \gls{cv} measurement is taken as main calibration target and other measurements are used to confirm that the compact model generalizes the device behavior. This quasi-static \gls{cv} measurement post wake-up is shown in Fig.~\ref{fig:iedm25:3}(a) together with the simulation from the compact model. The model captures the \gls{cv} curve closely, particularly for positive voltages, and is able to reproduce the asymmetry observed in the experimental curve. In particular, the asymmetry in the \gls{cv} peak amplitudes and the voltage bias are dictated by the interface charge density $N_{\text{tr,depl} \downarrow/\uparrow}$. These interface charges screen the polarization dipoles, i.e. the higher these charge densities are, the less the influence of the respective polarization dipoles on the depletion capacitance and therefore the \gls{cv} hysteresis. These parameters also reproduce the experimental P-V hysteresis loop (Fig.~\ref{fig:iedm25:3}(b)), as well as the switching kinetics for the erase state, illustrated in Fig.~\ref{fig:iedm25:3}(c). This result demonstrates consistency in the prediction of small-signal and large-signal behavior.
    
    Figure~\ref{fig:iedm25:4}(a) shows the capacitance versus cycling for high and low capacitance states at 0 V and 0.2 V read bias. The memory window increases only marginally at 0.2 V because the built-in field from the asymmetric stack creates a memory window at zero bias. This built-in field simplifies the read circuit design by eliminating the need for a bias voltage. Erase measurements are shown in Fig.~\ref{fig:iedm25:4}(b) and were obtained after 1\,k cycles through the initialization of the nvCap in the program state, followed by the application of erase pulses with magnitudes ranging from 2.5\,V to 4\,V in steps of 0.25\,V. The model accurately captures the gradual erase process at both voltages.

    \begin{figure}[t]
        \centering
        \includegraphics[]{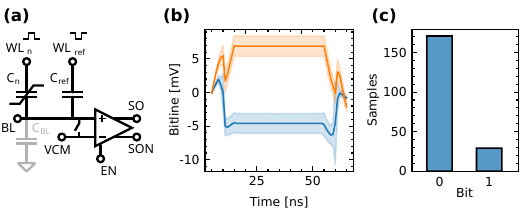}
        \caption{\textbf{(a)} In the digital memory macro, the currents of the nvCap and of the reference capacitor are summed up on the BL capacitance. \textbf{(a)} Monte Carlo circuit simulation in a commercial CMOS PDK with the nvCap compact model showing the voltage on the \gls{bl} during a 40\,ns read operation for LCS and HCS with 3 \,$\mathrm{\sigma}$.}
        \label{fig:iedm25:7}
    \end{figure}
    
    Furthermore, the capacitive memory window was verified at 0\,V and 200\,mV read voltages after every 1000 programming cycles of $\pm4$\,V/1\,ms (Fig.~\ref{fig:iedm25:5}(a)), and the variability of the \gls{lcs} and \gls{hcs} was extracted from the measurement. This variability was then included in the compact model by a variation of the electrode carrier density $N_\text{tr}$ with a standard deviation of $3.9 \cdot 10^{20} \text{cm}^{-3}$. This parameter, modeling the variability in the conductivity of the electrode at the interface, caused by oxidation of the electrode during both fabrication and operation, is a major contributor to the behavior and variability of ferroelectric devices \cite{alcala2023electrode}. Accordingly, we use this parameter to match mean and standard deviation of the measured memory windows. A Monte Carlo simulation of the \gls{cv} hysteresis with these variability parameters as shown in Fig.~\ref{fig:iedm25:5}(b), is used to select a suitable a read voltage for a robust read operation. The calibration reveals that, within the model, the variability of the capacitance is lowest at the peaks and increases the farther the bias is away from the peaks. The model accurately captures the erase process, as demonstrated in Fig.~\ref{fig:iedm25:5}(c), in conjunction with the experimental data. The Monte Carlo simulation also predicts a larger spread for HCS with a higher erase voltage, which is in accordance with the increase seen in Fig.~\ref{fig:iedm25:5}(c). 
    
    Endurance cycling results in the degradation of the \gls{nvcap}, exhibiting a comparable effect to that of device aging. Device endurance directly impacts circuit reliability. We now demonstrate how cycling-induced defect generation affects both the leakage current and the capacitive memory window, which both affect the read margin. This analysis enables circuit designers to set refresh requirements and error correction overhead. In the following we are focusing on two separate defect effects, as illustrated in Fig. \ref{fig:iedm25:2} (b). On the one hand, the bulk defect densities in the ferroelectric and interface layers $N_\text{tr,fe}$, which mainly affect the leakage current. On the other hand we investigate the interface defect densities for the up and down-polarized domains $N_{\text{tr,depl} \downarrow/\uparrow}$, which affect the electrostatics and hence the capacitance hysteresis. In order to extract the bulk defect densities $N_\text{tr,fe/int}$ from the devices, we used the Ginestra\texttrademark{} software to reproduce the experimental leakage current density data versus cycles for the erase state ~(Fig.~\ref{fig:iedm25:6}(a)). These TCAD-extracted defect densities then parameterize the compact model's TAT mechanism \mbox{(Eqs.~\ref{eq:tat0}--\ref{eq:tat9}).} The results indicate an increase in leakage, attributable to an increase of the defect densities $N_\text{tr,fe/int}$ resulting from cycling (Fig~\ref{fig:iedm25:6}(b)). The findings from the Ginestra~analysis are incorporated within the compact model through the implementation of a trap-assisted tunneling model \cite{larcher2012leakage, vandelli2011physical}. This linkage enables predictive modeling: changes in process conditions that alter the defect density can be simulated at the circuit level without re-extracting the full compact model. The degradation analysis requires the estimation of the trap density changes with cycling. The impact on small-signal \gls{cv} is illustrated in Fig.~\ref{fig:iedm25:8}(a) for various interface defect densities for the upwards polarization state $N_{\text{tr,depl}\uparrow}$. As previously observed, the non-volatile capacitance (at 0\,V) exhibits a substantial dependency on the interface trap density.
    
    \begin{figure}[t]
        \centering
        \includegraphics[]{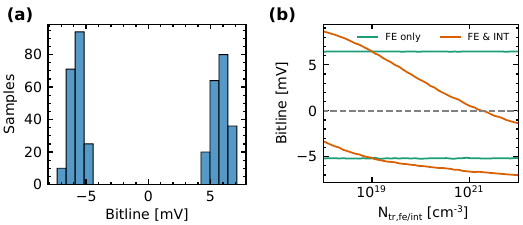}
        \caption{\textbf{(a)} Histogram of the \gls{bl} potential for LCS and HCS states in the Monte Carlo circuit simulation. \textbf{(b)} Circuit simulation showing \Gls{bl} signal degradation due to increasing bulk defects $N_\text{tr,fe/int}$ (in the FE only and in both FE and INT) during endurance cycling. As the leakage current~<~500\,mV is dominated by the interface, the FE defects and leakage alone have limited influence on the signal.}
        \label{fig:iedm25:7p5}
    \end{figure}
    
    The compact model is then employed to design and size a digital read-out macro, as in Fig.~\ref{fig:iedm25:7}. The sense amplifier (SA, Fig.~\ref{fig:iedm25:7}(a)) is a latch with a differential pair input \cite{fehlings2025reliability}. All following \gls{bl} and \gls{wl} voltages described here are relative to the common mode voltage that the \gls{bl} is pre-charged to. The circuit's operating principle entails the application of a 300\,mV read pulse, which induces a change in the charge on the positive input of the SA, proportional to the capacitance of the \gls{fecap}. This charge is added to the charge originating from a reference capacitor $C_\text{ref}$, normalizing the \gls{lcs} and \gls{hcs} voltages around 0. Consequently, this results in a change in voltage on the \gls{bl} (Fig.~\ref{fig:iedm25:8}(b)), which is pre-charged to a common-mode voltage $V_\text{cm}$ and is amplified by the latch. We performed Monte Carlo simulations including nvCap variability in the electrode carrier density $N_\text{depl}$ (Table \ref{tab:parameters}) as well as CMOS transistor mismatch and process variations sourced from the CMOS PDK. Figure~\ref{fig:iedm25:7p5}(a) demonstrates that a minimum sense margin of $\mathrm{\pm}$5\,mV can be achieved. The read yield, defined as the number of correct bit read-out divided by the number of Monte Carlo simulations carried out, for bit 0 is shown in Fig.~\ref{fig:iedm25:7}(c). An increase of variability in the electrode carrier density $N_{\text{depl}}$, or additional variability sources, would further decrease this sense margin. Here, our framework can enable the design process by suggesting which device variability source (e.g. $N_{\text{depl} \downarrow/\uparrow}$) to decrease, or how to adjust the read-out circuit or read voltage to accommodate for the sense margin degradation due to the device-level variability.
    
    The impact of the increase in trapped charges originating from interface defects on the up polarization state in the \gls{nvcap} capacitance hysteresis is illustrated in Fig.~\ref{fig:iedm25:8}(a). The trapped charges for the down polarization state are kept constant at $5 \cdot 10^{11}\,\text{cm}^{-2}$. The capacitive memory window increases with the number of interface defects $N_{\text{tr,depl}\uparrow}$, as well as the asymmetry between positive and negative voltage portions of the hysteresis. As the number of interface defects increases for the up polarization state, the capacitive memory window increases due to better polarization screening (Eq.~\ref{eq:cdepl_par}). Fig.~\ref{fig:iedm25:8}(b) shows the respective impact on the \gls{bl} voltage, where an increase in the aforementioned capacitive memory window translates to an increase in the \gls{bl} voltage for increasing defects. The combination of bulk defects in the ferroelectric and dielectric interface layer leads to a degradation of the signal on the \gls{bl}, as shown in Fig.~\ref{fig:iedm25:7p5}(b), due to higher leakage currents. Bulk defects in the ferroelectric layer alone do not significantly impact the read margin in our simulation. This is due to the leakage current being dominated by the interface contribution for voltages below 500\,mV in the compact model simulation, as the voltage drop over the interface dominates over the ferroelectric in that regime. This is in accordance with the experimental and \gls{tcad} simulation in Fig.~\ref{fig:iedm25:6}(a), where currents below 500\,mV show little dependence on the defect density in the ferroelectric $N_\text{tr,fe}$. An input offset of the sense amplifier of less than $\mathrm{\pm}$5\,mV is required when the bulk defect density increase above $\mathrm{10^{21}/\text{cm}^3}$.
    
    \begin{figure}[t]
        \centering
        \includegraphics[]{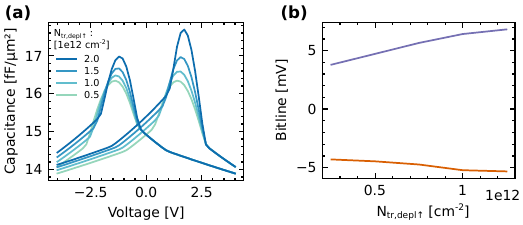}
        \caption{\textbf{(a)} Simulation of the compact model capacitance hysteresis as a function of interface defects for the up polarization state that act as trapping and polarization screening sites. \textbf{(b)} Circuit simulation of the BL signal as a function of the interface defects for the up polarization state. The interface modulates the polarization screening by trapped charges and hence the capacitive memory window (cf.~Fig.~\ref{fig:iedm25:5}(a)).}
        \label{fig:iedm25:8}
    \end{figure}
    
    Based on the analysis done with our framework, we derive the following device and circuit design guidelines: Due to the low capacitive memory window, particularly for devices with scaled area, there is a strong tradeoff in device area, bit line capacitance, read voltage and sense amplifier. In particular for larger arrays and under the consideration of noise, device and circuit have to undergo in-depth co-design. To increase the available capacitive memory window, our framework suggests a strong asymmetry between interface charges ($N_{\text{tr,depl} \downarrow/\uparrow}$) for the up and down polarization directions. This comes, however, at the cost of an increase built-in bias, potentially increasing program and erase voltages, and could pose reliability challenges. The variability in the capacitive memory window and read-out yield shows a strong dependence on the electrode screening properties, with a variation in the carrier density $N_\text{depl}$ showing large read signal variability. An optimization of the yield then requires strong control of the electrode interface on the device side and offset compensation below a few millivolts for the sense amplifier. Bulk traps, especially for the dielectric \gls{al2o3} layer ($N_\text{tr,int}$), contribute severely to the leakage current, and an increase during the devices life-cycle shows a decrease in the bit line signal. The device therefore needs to be optimized to limit defects that participate in trap-assisted tunneling to limit the decrease in bit line signal during aging.

    \begin{table}
    \caption{Model parameters}
    \label{table}
    \setlength{\tabcolsep}{3pt}
    \begin{tabular}{p{35pt} p{65pt} p{125pt}}
    \hline
    Parameter & Value & Description \\
    \hline
    $\alpha_\text{polar}$ & 0.6 & Fraction of ferroelectric domains\\
    $\varepsilon_\text{fe}$ & 70 & Permittivity of ferroelectric domains\\
    $\varepsilon_\text{de}$ & 20 & Permittivity of dielectric domains\\
    $\varepsilon_\text{depl}$ & 2.2 & Polarizability electrode interface\\
    $P_\text{s}$ & 20 \textmu C/cm\textsuperscript{2} & Saturation polarization\\
    $N_\text{depl}$ & 1.2 $\cdot$ 10\textsuperscript{22} cm\textsuperscript{-3} & Carrier density electrode interface\\
    $\sigma(N_\text{depl})$ & 3.9 $\cdot$ 10\textsuperscript{20} cm\textsuperscript{-3}  & Std. dev. $N_\text{depl}$ (Monte Carlo only)\\
    $N_{\text{tr,depl}\uparrow}$ & 0 & Screening charge density P$_{\uparrow}$ \\
    $N_{\text{tr,depl}\downarrow}$ & 7.5 $\cdot$ 10\textsuperscript{13} cm\textsuperscript{-2} & Screening charge density P$_{\downarrow}$ \\
    $\varphi_\text{b,fe}$ & 2 eV & FE layer conduction band offset\\
    $\varphi_\text{b,int}$ & 2.7 eV & INT layer conduction band offset\\
    $N_\text{tr,fe}$ & 1.5 $\cdot$ 10\textsuperscript{19}  cm\textsuperscript{-3} & Trap density in the FE\\
    $m^*_\text{fe}$ & 0.4 & Electron effective mass\\
    $W_\text{tr,t}$ & 2.1 eV & Thermal ionization energy \\
    $W_\text{tr,rel}$ & 1.19 eV & Relaxation energy \\
    $x_\text{c}$& 5 nm & Tunneling characteristic length \\
    $c_0N_\text{c}$& 1 $\cdot$ 10\textsuperscript{-8}  s\textsuperscript{-1} & Trap-specific constant \\
    \hline
    \end{tabular}
    \label{tab:parameters}
    \end{table}

\section{Conclusion}
    We propose a material-device-circuit co-design framework for developing a non-destructive read-out of \gls{nvcap}. This framework is based on experimental characterization of a bi-layer \gls{nvcap} device, kinetic Monte Carlo-based \gls{tcad} simulations and a physics-based compact model. The proposed compact model has the capacity to replicate the behavior of the \gls{nvcap} including the modeling of defect contributions to the capacitance hysteresis and leakage currents. We conclude that interface defects can increase the sense margin due to better polarization screening, while bulk defects, particularly in the Al\textsubscript{2}O\textsubscript{3} interface layer, lower the sense margin due to increased leakage currents. Our framework further allows to extract device variability to model process variability and mismatch. These results are then applied to a memory circuit where sense margins are extracted based on variability and defect parameters. The defect-based modeling approach enables circuit designs that compensate for reliability and aging. This capability is essential for advanced storage and compute-in-memory applications where array size and endurance requirements challenge conventional design margins. By linking process and cycling-induced defects to circuit-level performance, this framework supports technology-circuit co-optimization from early process development through memory array design.

\section*{Acknowledgement}

The authors thank Gaurav Thareja and Luca Larcher from Applied Materials for their mentorship and access to the Ginestra\texttrademark~proprietary simulation platform.\\

This work was supported by the European Research Council (ERC) through the European's Union Horizon Europe Research and Innovation Programme under Grant Agreement No 101042585. Views and opinions expressed are however those of the authors only and do not necessarily reflect those of the European Union or the European Research Council. Neither the European Union nor the granting authority can be held responsible for them. The University of Groningen would like to acknowledge the financial support of the CogniGron research center and the Ubbo Emmius Funds. IIT Bombay acknowledges funding from DST and MeitY through NNETRA project, iHUB DivyaSampark - IIT Roorkee. Support of IITBNF staff is acknowleged.



\normalsize
\bibliography{references}


\end{document}

%% file: glossary.tex
\newacronym{1t1c}{1T1C}{one transistor, one capacitor}
\newacronym{1t1r}{1T1R}{one transistor, one resistor}
\newacronym{2t1c}{2T1C}{two transistor, one capacitor}
\newacronym{adc}{ADC}{analog to digital converter}
\newacronym{ae}{AE}{active electrode}
\newacronym{aer}{AER}{address-event representation}
\newacronym{ai}{AI}{artificial intelligence}
\newacronym{al2o3}{Al\textsubscript{2}O\textsubscript{3}}{alumina}
\newacronym{ald}{ALD}{atomic layer deposition}
\newacronym{asic}{ASIC}{application-specific integrated circuit}
\newacronym{be}{BE}{bottom electrode}
\newacronym{beol}{BEOL}{back-end-of-line}
\newacronym{bl}{BL}{bit line}
\newacronym{bptt}{BPTT}{backpropagation through time}
\newacronym{bruno}{BRUNO}{backpropagation running undersampled for novel device optimization}
\newacronym{bsim}{BSIM}{Berkeley short-channel IGFET model}
\newacronym{c2c}{C2C}{cycle-to-cycle}
\newacronym{cam}{CAM}{content-addressable memory}
\newacronym{cc}{CC}{current conveyor}
\newacronym{cim}{CIM}{compute-in-memory}
\newacronym{cmos}{CMOS}{complementary metal oxide semiconductor}
\newacronym{cv}{CV}{capacitance-voltage}
\newacronym{dac}{DAC}{digital to analog converter}
\newacronym{dhm}{DHM}{dynamic hysteresis measurement}
\newacronym{dnn}{DNN}{deep neural network}
\newacronym{dpi}{DPI}{differential pair integrator}
\newacronym{dram}{DRAM}{dynamic random access memory}
\newacronym{dtco}{DTCO}{design-technology co-optimization}
\newacronym{dut}{DUT}{device under test}
\newacronym{eda}{EDA}{electronic design automation}
\newacronym{envm}{eNVM}{embedded non-volatile memory}
\newacronym{fecap}{FeCap}{ferroelectric capacitor}
\newacronym{felif}{FeLIF}{FeCap-based leaky-integrate-and-fire}
\newacronym{fefet}{FeFET}{ferroelectric field-effect transistor}
\newacronym{fet}{FET}{field-effect transistor}
\newacronym{feol}{FEOL}{front-end-of-line}
\newacronym{fgfet}{FGFET}{floating-gate field-effect transistor}
\newacronym{forc}{FORC}{first-order reversal curve}
\newacronym{fram}{FRAM}{ferroelectric random-access memory}
\newacronym{ftj}{FTJ}{ferroelectric tunnel junction}
\newacronym{gpu}{GPU}{graphics processing unit}
\newacronym{hcs}{HCS}{high-capacitance state}
\newacronym{hfo2}{HfO\textsubscript{2}}{hafnia}
\newacronym{hkmg}{HKMG}{high-k metal gate}
\newacronym{hpo}{HPO}{hyperparameter optimization}
\newacronym{hrs}{HRS}{high resistive state}
\newacronym{hzo}{HZO}{hafnium zirconium oxide (Hf\textsubscript{0.5}Zr\textsubscript{0.5}O\textsubscript{2})}
\newacronym{ic}{IC}{integrated circuit}
\newacronym{if}{I\&F}{integrate-and-fire}
\newacronym{imc}{IMC}{in-memory computing}
\newacronym{ip}{IP}{intellectual property}
\newacronym{iv}{IV}{current-voltage}
\newacronym{jart}{JART}{Jülich Aachen resistive switching tool}
\newacronym{kai}{KAI}{Kolmogorov-Avrami-Ishibashi}
\newacronym{lcs}{LCS}{low-capacitance state}
\newacronym{li}{LI}{leaky integrator}
\newacronym{lif}{LIF}{leaky integrate-and-fire}
\newacronym{lrs}{LRS}{low resistive state}
\newacronym{lut}{LUT}{look-up table}
\newacronym{mac}{MAC}{multiply and accumulate}
\newacronym{mc}{MC}{monte carlo}
\newacronym{mfim}{MFIM}{metal-ferroelectric-insulator-metal}
\newacronym{mfm}{MFM}{metal-ferroelectric-metal}
\newacronym{mfis}{MFIS}{metal-ferroelectric-insulator-semiconductor}
\newacronym{ml}{ML}{machine learning}
\newacronym{mos}{MOS}{metal-oxide-semiconductor}
\newacronym{mosfet}{MOSFET}{metal-oxide-semiconductor field-effect transistor}
\newacronym{mram}{MRAM}{magnetic random-access memory}
\newacronym{nand}{NAND}{nand-gate memory}
\newacronym{nbox}{NbO$\mathrm{_x}$}{NbO$\mathrm{_x}$}
\newacronym{nmos}{NMOS}{n-type metal-oxide-semiconductor}
\newacronym{nor}{NOR}{nor-gate memory}
\newacronym{nvcap}{nvCap}{non-volatile capacitor}
\newacronym{nvdram}{NVDRAM}{non-volatile dynamic random access memory}
\newacronym{nvm}{NVM}{non-volatile memory}
\newacronym{oe}{OE}{ohmic electrode}
\newacronym{ota}{OTA}{operational transconductance amplifier}
\newacronym{pcm}{PCM}{phase-change memory}
\newacronym{pdk}{PDK}{process design kit}
\newacronym{pf}{PF}{poole-frenkel}
\newacronym{pl}{PL}{plate line}
\newacronym{pmos}{PMOS}{p-type metal-oxide-semiconductor}
\newacronym{ppa}{PPA}{power performance area}
\newacronym{pund}{PUND}{positive-up-negative-down}
\newacronym{pv}{PV}{polarization-voltage}
\newacronym{pvt}{PVT}{process voltage temperature}
\newacronym{pzt}{PZT}{Pb[Zr${\mathrm{_x}}$Ti$\mathrm{_{1-x}}$]O$_3$}
\newacronym{qat}{QAT}{quantization aware training}
\newacronym{rram}{RRAM}{resistive switching non-volatile devices}
\newacronym{rsnn}{rSNN}{recurrent spiking neural network}
\newacronym{scm}{SCM}{storage-class memory}
\newacronym{sio2}{SiO\textsubscript{2}}{SiO\textsubscript{2}}
\newacronym{sl}{SL}{source line}
\newacronym{snn}{SNN}{spiking neural network}
\newacronym{spice}{SPICE}{simulation program with integrated circuit emphasis}
\newacronym{sram}{SRAM}{static random access memory}
\newacronym{stc}{STC}{synaptic tagging and capture}
\newacronym{stco}{STCO}{system-technology co-optimization}
\newacronym{ste}{STE}{straight-through estimator}
\newacronym{spc}{SPC}{statistical process control}
\newacronym{tat}{TAT}{trap-assisted tunneling}
\newacronym{tcad}{TCAD}{technology computer-aided design}
\newacronym{tddb}{TDDB}{time-dependent dielectric breakdown}
\newacronym{te}{TE}{top electrode}
\newacronym{vcm}{VCM}{valence change mechanism}
\newacronym{wl}{WL}{word line}
\newacronym{wm}{WM}{working memory}
\newacronym{wta}{WTA}{winner-take-all}